\documentclass[preprint,12pt]{elsarticle}

\usepackage{amssymb}
\usepackage{amsmath}
\usepackage{xcolor}

\journal{Physica A}

\begin{document}

\begin{frontmatter}



\title{Negotiation Problem}


\author[add1]{Izat B. Baybusinov}
\author[add1]{Enrico Maria Fenoaltea}
\author[add1]{Yi-Cheng Zhang\corref{mycorrespondingauthor}}
\cortext[mycorrespondingauthor]{Corresponding author}
\ead{yi-cheng.zhang@unifr.ch}

\address[add1]{Physics Department, University of Fribourg Chemin du Mus\'ee 3, 1700                                Fribourg, Switzerland}


\begin{abstract}
    We propose and solve a negotiation model of multiple players facing many alternative solutions. The model can be generalized to many relevant circumstances where stakeholders' interests partially overlap and partially oppose. We also show that the model can be mapped into the well-known directed percolation and directed polymers problems. Moreover, many statistical mechanics tools, such as the Replica method, can be fruitfully employed. Studying our negotiation model can enlighten the links between social-economic phenomena and traditional statistical mechanics and help to develop new perspectives and tools in the fertile interdisciplinary field. 
\end{abstract}


\begin{highlights}
\item We propose and solve a non-zero sum game of negotiation.
\item We study its Nash equilibrium and ground state.
\item We employ both a combinatorial statistical approach and standard statistical mechanics tools.
\item The model can be mapped into two well-known physics problems: directed percolation and directed polymers.
\end{highlights}


\end{frontmatter}


\section{Introduction}
Often in social and economic processes we must reach an agreement with other people. We may have partially overlapping interests and conflicts which, to settle, we must negotiate.
    To study such behaviour, we propose a simple model of negotiation which turns out to belong to a well-known class of physics models, ranging from spin-glasses to random matching, and other models in physics and beyond \cite{sherrington1975solvable,mezard1985replicas, mezard1986replica,  mezard1986mean,fenoaltea2021stable}.
    
    In principle, our problem follows the same philosophy of the bargaining problem \cite{Nash,rubinstein1982perfect}. The latter is a non-cooperative game where agents compete to maximize their utilities. Each agent is fully rational and knows the preferences of all the others. The main goal of the bargaining problem is studying the conditions for a Nash equilibrium, i.e. a situation where, for each agent, it is not convenient to change strategy \cite{standardgame}. 
    In this approach, each agent can choose among infinite possible strategies and evaluate the opponents' one before making a move.
    
    By contrast, we consider bounded-rational players that cannot span across all possible choices, but are limited to few, depending on their current knowledge \cite{bounded}.
    Indeed, here we want to model a more realistic scenario where people, during the negotiation, have limited time to make choices and must find a compromise between time spent and expected utility. We consider a non-zero sum game: the utility of each player is independent on all the others, rather than being  anti-correlated \cite{rubinstein1982perfect}. This means that people have different tastes and preferences that are not always conflicting. This approach is inspired by the Stable Marriage Problem \cite{fenoaltea2021stable}, where players have independent preference-lists.   
    
    
    In this paper, we will study analytically and numerically the statistical properties of our negotiation model. The rest of the paper is organized as follows:  In section 2, we formally describe the model with 2 players, studying its Nash-equilibrium solution and its ground state, i.e. the global best solution; In section 3 we generalize our model to more than 2 players and show its connection to other known physics models; Finally, in the last section, we expose our conclusions.
    

  \section{Negotiation Model}

	Let us consider the case of two players, A and B, who must find an agreement among $N$ alternatives. Each player ranks these alternatives according to his preference, from the best to the worst. The preference-lists are randomly drawn, and each alternative is acceptable to both players, even the worst, with descending benefits.

	We assume that players propose alternatives, in sequence, from their most favourite down to the least favourite, by taking alternate turns. If player A is the first mover (i.e. the first who proposes), then at time $t=1$ he proposes his best choice to player B, and player B can either accept it, ending the negotiation, or reject it. In the latter case, the negotiation continues, and it will be player B's turn to propose his first-ranked alternative; if also player A rejects, then the negotiation proceeds to $t=2$, and player A must propose his second-best choice, as his first one was already rejected. The process continues with the same logic until a proposal is accepted. 
	This happens when, at a given time $t$, player A (or B) proposes his next alternative to player B (or A), to whom it happens to be better (for the first time) than the alternatives that he would propose next. So player B (or A) accepts and the negotiation ends with a solution. Note that, by construction, an alternative is accepted only if it is the first alternative to have been proposed by both players. An example of negotiation is shown in Fig.1. 
	
	Let us denote the time to reach an agreement by $t_f$, the finally agreed alternative by $f$, and its respective ranking in the two players' preference-lists by $\epsilon_{{\scriptscriptstyle A}}$ and $\epsilon_{{\scriptscriptstyle B}}$. Analogously to the matching problem \cite{fenoaltea2021stable}, we use the term energy to describe ranking: alternative $i$ has an assigned energy $\epsilon_{i,{\scriptscriptstyle A}}$ for A, and $\epsilon_{i,{\scriptscriptstyle B}}$ for B. For simplicity, energies assume integer values, i.e. $\epsilon_i=1,2,...,N$.
	\begin{figure}[!h]
	    \centering
	    \includegraphics{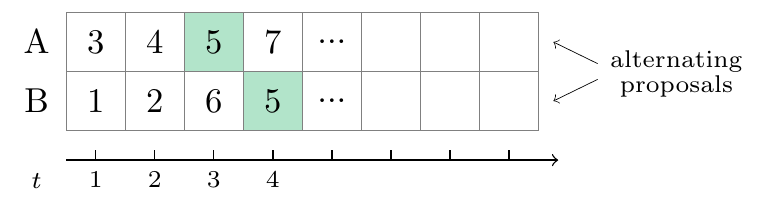}
	    \caption{Figurative representation of the negotiation process: the rows represent the alternative proposed by player A (top row) and by player B (bottom row). The proposals are made from left to right and the alternatives are labeled with natural numbers from 1 to N (of course, these have nothing to do with the energies). Here only the four best alternatives in the preference-lists of the two players are shown: in particular, the first choice of player A is the alternative 3, while the fourth is the alternative 7; for player B, instead, the best choice is the alternative 1, and the fourth-best is the alternative 5. Hence, player A starts by proposing the alternative 3, then player B proposes the alternative 1 and so on. The process stops when player A accepts the alternative 5 proposed by player B at time $t_f=4$. Indeed, for player A, his subsequent proposals would have been worse than alternative 5. Note that the alternative 5 is the only one proposed by both players until that moment and, actually, this is the condition for an alternative to be accepted. In this example the energies are $\epsilon_{5,A}=\epsilon_A=3$ and $\epsilon_{5,B}=\epsilon_B=4$.}
	    \label{fig:exlfigure}
	\end{figure}

	\subsection{Negotiation Solution}
	 In the negotiation solution we are interested in the energy of both players,  $\epsilon_{{\scriptscriptstyle A}}$ and $\epsilon_{{\scriptscriptstyle B}}$, and in  the energy gap, $d=|\epsilon_{{\scriptscriptstyle A}}-\epsilon_{{\scriptscriptstyle B}}|$, which tells how fair the solution is. We evaluate these quantities by sample averaging them over all configurations of the preference-lists, and we shall denote the sample average of an observable $o$ by $\overline{o}$.
	
	Firstly we compute the probability $P(\epsilon_{{\scriptscriptstyle A}},\epsilon_{{\scriptscriptstyle B}})$ of finding the negotiation solution with energies $\epsilon_{{\scriptscriptstyle A}}$ and $\epsilon_{{\scriptscriptstyle B}}$.
    
    When $\epsilon_{{\scriptscriptstyle A}} > \epsilon_{{\scriptscriptstyle B}}$, the alternative proposed by player A at time $t_f=\epsilon_{{\scriptscriptstyle A}}$ is accepted, while all previous proposals up to time $t_f-1$ are rejected.
    The number of times this situation occurs is given by the number of ways of choosing $ t_f-1 $ alternatives out of the $ t_f- \epsilon_A $ alternatives worse than $ \epsilon_A $, i.e. $\binom{N-\epsilon_{{\scriptscriptstyle A}}^{} }{\epsilon_{{\! \scriptscriptstyle A}}^{}-1}(\epsilon_{{\scriptscriptstyle A}}^{}-1)!$. 
     Since there are $\binom{N}{\,\epsilon_{{\!\scriptscriptstyle A}}}\epsilon_{{\scriptscriptstyle A}}^{}!$ ways of choosing $\epsilon_{{\scriptscriptstyle A}}$ elements out of $N$, the probability is
    \begin{equation}
	    P(\epsilon_{{\scriptscriptstyle A}}, \epsilon_{{\scriptscriptstyle B}}) =
        \frac{\binom{N-\epsilon_{{\!\scriptscriptstyle A}}}{\epsilon_{{\scriptscriptstyle A}}-1}}{\binom{N}{\epsilon_{{\!\scriptscriptstyle A}}^{}}\epsilon_{{\scriptscriptstyle A}}}, \quad   \epsilon_{{\scriptscriptstyle A}} > \epsilon_{{\scriptscriptstyle B}} \;. 
	\end{equation}
	
	Expanding the binomial coefficients, we can rewrite this as follows:
	\begin{equation}
	 P(\epsilon_{{\scriptscriptstyle A}}, \epsilon_{{\scriptscriptstyle B}}) = \frac{1}{N-\epsilon_{{\scriptscriptstyle A}}+1}\prod_{i=1}^{\epsilon_{{\scriptscriptstyle A}}-1}\left(1-\frac{\epsilon_{{\scriptscriptstyle A}}}{N-i+1}\right) . 
	\end{equation}
	In this equation, the first factor is the probability that player A accepts a proposal at time $t_f=\epsilon_{{\scriptscriptstyle A}}$, while the product represents the probability that player A rejects all the proposals before $t=t_f$.\\
	By symmetry, the case $\epsilon_{{\scriptscriptstyle B}} > \epsilon_{{\scriptscriptstyle A}}$ is obtained by exchanging $A$ and $B$. 
	Combining both cases, for large $N$ we obtain
	\begin{equation}
	    P(\epsilon_{{\scriptscriptstyle A}}, \epsilon_{{\scriptscriptstyle B}}) \approx
        \frac{1}{N}  \exp{\left\{-\frac{t_f^2}{N}\right\}} \;,
	\end{equation}
	where $t_f=\max\left(\epsilon_{{\scriptscriptstyle A}},\epsilon_{{\scriptscriptstyle B}}\right)$.
	From this, we can compute the sample averages:
    \begin{align}
    \overline{\epsilon_{{\scriptscriptstyle A}}}=\overline{\epsilon_{{\scriptscriptstyle B}}}  =\frac{3\sqrt{\pi}}{8}\sqrt{N} \approx 0.665 \sqrt{N}\;,
    \end{align}
    \begin{equation}
    \overline{d}=\frac{\sqrt{\pi}}{4}\sqrt{N} \approx 0.443\sqrt{N} \;.
    \end{equation}
     The probability that the two players have the same energy is $P(\epsilon_{{\scriptscriptstyle A}}=\epsilon_{{\scriptscriptstyle B}})=\sqrt{\pi}/2\sqrt{N}$, showing that such cases are rare when $N\to \infty$.

     We also calculate the players' energy distribution, i.e. $P(\epsilon_{{\scriptscriptstyle A}})=\sum_{\epsilon_{\scriptscriptstyle B}} P(\epsilon_{{\scriptscriptstyle A}},\epsilon_{{\scriptscriptstyle B}}) $. From (3) we have
    \begin{equation}
        \sum_{\epsilon_{\scriptscriptstyle B}} P(\epsilon_{{\scriptscriptstyle A}},\epsilon_{{\scriptscriptstyle B}}) \approx
        \frac{\displaystyle{\epsilon}_{{\! \scriptscriptstyle A}}}{N}e^{-
        \epsilon_{{\! \scriptscriptstyle A}}^{ 2}/{N}}
        +
        \int\limits_{\epsilon_{\scriptscriptstyle A}+1}^{N} \frac{d\epsilon_{{\scriptscriptstyle B}}}{N}\, 
        e^{-\epsilon_{{\!\scriptscriptstyle B}}^2/N} \;.
    \end{equation}
    
    \begin{figure}
        \centering
        \includegraphics{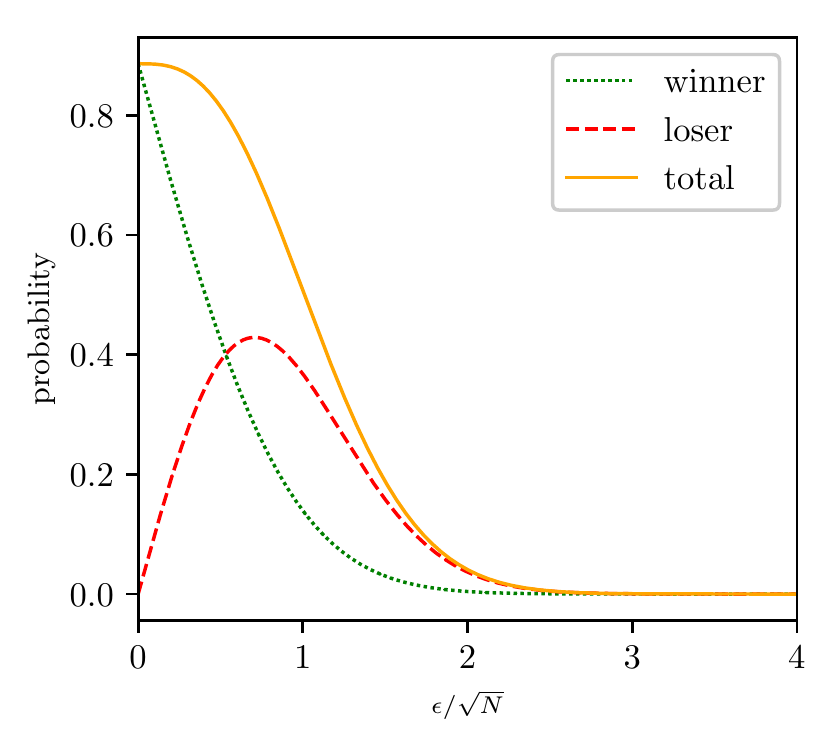}
        \caption{Probability distribution of the total energy compared with that of both the winner (the player with lower energy) and the loser (the player with higher energy). }
        \label{fig:winlose}
    \end{figure}
    
    In the right hand side of (6), the first term is the probability of obtaining a higher energy than the other player, and the second is the probability of obtaining a lower energy (Fig. 2).

    
    
    Note that, in the negotiation model, all rules are fair except for the first move. Indeed, the player who moves first, on average, loses more (i.e. he obtains higher energy than the opponent). One can show that the stopping time $t_f$ is independent of the first mover, therefore the solution depends on the first mover whenever both proposals of player A and player B at time $t_f$ can be accepted by the other player. This happens with probability $t_f/2N$, and the energy difference is, on average, $t_f/2$. It follows that the average first mover disadvantage is given by:
    \begin{equation}
        \overline{\epsilon_A-\epsilon_B}=\overline{\frac{t_f}{2N}\frac{t_f}{2} }=\frac{1}{4}+O(N^{-1/2})
    \end{equation}
    However, this is a negligible effect when $N\rightarrow\infty$.
    
    Nevertheless, the first-mover disadvantage provides insight into the negotiation model. Whoever moves first, or by extension moves more steps, will bear additional costs. This is because whoever exposes more information, shows more cards so to speak, is at a disadvantage. Therefore rational and selfish players will show as little and as late as possible their cards. 
	This, \textit{a posteriori}, justifies that our strategy is optimal for each player.

 \subsection{Ground State Solution}

    Now, from physicists' point of view, the most interesting is the Ground State, i.e. the solution that minimizes the total energy.
	 
	The minimal energy is usually found by defining a Hamiltonian, but we shall follow the probabilistic approach first, then compare it to the traditional Hamiltonian approach.
	
    The ground state is defined as
	\begin{align}
	    \epsilon_{\scriptscriptstyle 0}= \min\limits_{i=1,..,N} \{\epsilon_{i,{\scriptscriptstyle A}} + \epsilon_{i,{\scriptscriptstyle B}}\}=
	    \epsilon_{\scriptscriptstyle 0,A}+\epsilon_{\scriptscriptstyle 0,B}\;.
	\end{align}
	To find it, we compute the cumulative distribution function of $\epsilon_{{\scriptscriptstyle 0}}$:
	\begin{align}
	    P(\epsilon_{{\scriptscriptstyle 0}} \geq l) = \prod\limits_{i=1}^N P(\epsilon_{i,{\scriptscriptstyle A}} + \epsilon_{i,{\scriptscriptstyle B}} \geq l) \;.
	\end{align}
	We set $\epsilon_{i,{\scriptscriptstyle A}}=i$ and $\epsilon_{i,{\scriptscriptstyle B}}=\pi_N(i)$, where $\pi_N$ is a random permutation, 
	obtaining $P(\epsilon_{i,{\scriptscriptstyle B}}\geq l-i)=\frac{N-l+1}{N-i+1}$ for $i<l$. So (9) can be rewritten as 
	\begin{align}
	    P(\epsilon_{{\scriptscriptstyle 0}} \geq l) = \prod\limits_{i=1}^{l-1} \frac{N-l+1}{N-i+1}=
	    \prod_{i=1}^{l-1}\left(1-\frac{l-i}{N-i+1}\right) \;,
	\end{align}
	which for $N\to\infty$ converges to:
	\begin{align}
	    P(\epsilon_{{\scriptscriptstyle 0}} \geq l) \approx \exp{\left(-\frac{l(l-1)}{2N} \right)}\;.
	\end{align}
	From (11) we can compute the average as $\overline{\epsilon_{{\scriptscriptstyle 0}} \mathstrut}=\sum\limits_{l=1}^{N+1}P(\epsilon_{{\scriptscriptstyle 0}}\geq l)$, so the average energy per player in the ground state, $\overline{\epsilon_{{\scriptscriptstyle 0}}  }/2 =\overline{\epsilon_{\scriptscriptstyle 0,A} }=\overline{ \epsilon_{\scriptscriptstyle 0,B }}$, is
	\begin{align}
	    \overline{\epsilon_{\scriptscriptstyle 0,A} \mathstrut}=\overline{ \epsilon_{\scriptscriptstyle 0,B }\mathstrut}= \sqrt{\frac{\pi}{8}}\sqrt{N} \approx 0.627 \sqrt{N} \;.
	\end{align}
    The average energy gap is obtained by averaging all the possible $\epsilon_{\scriptscriptstyle 0,A}$ and $\epsilon_{\scriptscriptstyle 0,B}$  for a given energy value of the ground state. Since the energies can take values from $1$ to $\epsilon_{{\scriptscriptstyle 0}}-1$, the average gap is equal to the average energy per player (true only if sample averaged):
    \begin{align}
	    \overline{|\epsilon_{\scriptscriptstyle 0,A}-\epsilon_{\scriptscriptstyle 0,B}|}= \overline{\epsilon_{\scriptscriptstyle 0,A} \mathstrut}=\overline{ \epsilon_{\scriptscriptstyle 0,B }\mathstrut} \;,
	\end{align}

  The ground state energy in (12) is about $6\%$ lower than that of the negotiation solution in (4).
  On the other hand, comparing (13) with (5), we see that the energy gap in the ground state is about $42\%$ larger than that in the negotiation solution. Indeed, the ground state has no fairness constraint (as players have no decision-making process), and, though the total energy is minimal, the energy gap is larger. Thus, a lack of equality is the price to pay for reducing the total energy.
  
  Incidentally, the two solutions often are identical. As it turns out, with $\frac{\pi}{4}\approx78\%$ of probability, the negotiation finds the ground state exactly, when $N \to \infty$.
  
  To see it, we compute the probability $P(\epsilon_{{\scriptscriptstyle 0}}=\epsilon_{{\scriptscriptstyle A}} + \epsilon_{{\scriptscriptstyle B}})$ that the two solutions coincide. In the large $N$ limit, we have
	\begin{align}
	    P(\epsilon_{{\scriptscriptstyle 0}}=\epsilon_{{\scriptscriptstyle A}}+\epsilon_{{\scriptscriptstyle B}}) 
	    \approx 
	    \frac{1}{N}
	    \exp\left\{\frac{-\epsilon_{{\!\scriptscriptstyle A}}^2-\epsilon_{{\scriptscriptstyle B}}^2+\epsilon_{{\!\scriptscriptstyle A}}+\epsilon_{{\scriptscriptstyle B}}}{N}\right\}\;.
	\end{align}
	
    Summing over all possible energies, the probability that the ground state and the negotiation solution overlap is:
	\begin{align}
	     P(\text{overlap})=\sum\limits_{\epsilon_{{\scriptscriptstyle A}},\epsilon_{{\scriptscriptstyle B}}}P(\epsilon_{{\scriptscriptstyle 0}}=\epsilon_{{\scriptscriptstyle A}}+\epsilon_{{\scriptscriptstyle B}})=\frac{\pi}{4}\;.
	     	\end{align}

	\section{Negotiation model for $m$ players}
	It is interesting to generalize the negotiation model to more players. Not only it is a natural generalization like from Ising to Potts model \cite{ising,potts}, but it is also of wider applications. Indeed, seeking an agreement between several parties is common in diplomatic and business negotiations: as an example, European Union negotiations take place with 26 countries and each agreement must be accepted by all.
	
	\subsection{Negotiation Solution}
	Consider $m$ players that must find an agreement among $N$ alternatives, but each has a distinct preference-list. Again, each player takes turns to propose and all the others may agree or reject, as in the two-players model. Note that a partial agreement is not enough: even if only one player rejects a proposal, the negotiation must continue. 
	
	Let us first study the case of three players, $A$, $B$ and $C$. Analogously to the two-player model, we compute the probability $P(\epsilon_{{\scriptscriptstyle A}},\epsilon_{{\scriptscriptstyle B}},\epsilon_{{\scriptscriptstyle C}})$ of finding the negotiation solution with energies $\epsilon_{{\scriptscriptstyle A}}$, $\epsilon_{{\scriptscriptstyle B}}$ and $\epsilon_{{\scriptscriptstyle C}}$.
	To simplify, we divide the energies by $N$ and we consider the case  $\epsilon_{{\scriptscriptstyle A}} = \max\left( \epsilon_{\scriptscriptstyle A},\epsilon_{\scriptscriptstyle B},\epsilon_{\scriptscriptstyle C} \right)$.
	Hence, $P(\epsilon_{{\scriptscriptstyle A}},\epsilon_{{\scriptscriptstyle B}},\epsilon_{{\scriptscriptstyle C}})$ is equivalent to the probability that both player $B$ and player $C$ accept the proposal of player $A$ after $N\epsilon_{{\scriptscriptstyle A}}-1$ rejections. In the $N\to \infty$ limit, we have 
	\begin{equation}
    \begin{split}
    P(\epsilon_{{\scriptscriptstyle A}},\epsilon_{{\scriptscriptstyle B}},\epsilon_{{\scriptscriptstyle C}}) \approx 
    N\exp\left\{- N \epsilon_{{\scriptscriptstyle A}}^{3}\right\} \;.
    \end{split}
    \end{equation}
    
	In the general case of $m$ players, similar computations lead to 
    \begin{equation}
    \begin{split}
    P(\epsilon_1,...,\epsilon_m) \approx 
    N\exp\left\{-N t_f^{m}\right\} \;,
    \end{split}
     \end{equation}
    where $\epsilon_i$ denotes the energy of player $i$ and $t_f=\max\left(\epsilon_1,...,\epsilon_m\right)$.
    
    From (17) it follows that the average energy per player is:
    \begin{equation}  
    \begin{split}
    \overline{\epsilon}\equiv \frac{1}{m}\overline{\sum_{i=1}^m\epsilon_i}&=
    \frac{m+1}{2m}\Gamma\left(1+\frac{1}{m}\right)
    N^{ - {1}/{m}} \;.
    \end{split} 
    \end{equation}
    As expected, the larger $ m $, the higher the average energy per player, as there are more constraints in the negotiation. However, one can easily show that a larger $m$ implies a smaller relative energy gap between the players.
    
    \subsection{Ground State Solution}
    We now compute the ground state $\epsilon_{{\scriptscriptstyle 0}}$ for $m$ players and we consider only the large $N$ limit, so that we can approximate the preference-lists by $N$ random and independent variables in the interval $[0,1]$. Hence, the extreme value distribution of $N$ random variables $\epsilon_i$ is
    \begin{equation}
    P(\epsilon_{{\scriptscriptstyle 0}}>l)=\left(1-F(l)\right)^N \;, 
    \end{equation}
    where, in our case, $F(x)$ is the cumulative distribution function of the sum of $m$ random variables. Since the largest contribution to (19) is for $l\sim0$, the cumulative distribution can be approximated by $F(l)\sim 
    \frac{l^{m}}{m!}$, obtaining a Weibull distribution \cite{galambos1978asymptotic}:
    \begin{equation}  
    P(\epsilon_{{\scriptscriptstyle 0}}>l)=\exp \left\{-N{l^{m}}/{m!}\right\} \;.
    \end{equation}
    From (20), the average energy per player in the ground state is:
    \begin{equation}
    \frac{\overline{\epsilon_{{\scriptscriptstyle 0}}}}{m}=
    \int_0^\infty dl P(\epsilon_{{\scriptscriptstyle 0}}>l)=
    \frac{m!^{1/m}}{m}\Gamma\left(1+\frac{1}{m}\right) N^{-1/m}\;.
    \end{equation}
    
    \begin{figure}[!ht]
    \centering
    \hspace*{-0.7cm}

    \includegraphics[scale=1]{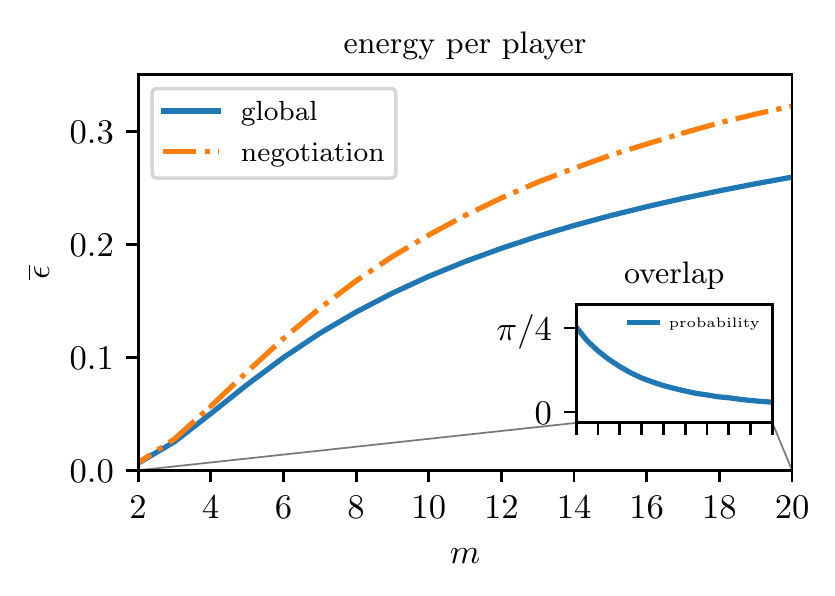}

    \caption{
    Average energy per player against the number of players $m$, both for negotiation and ground state solutions, with N=10000. In the inset: the probability that negotiation and ground state solutions coincide against the number of player $m$. As shown in (15), with two players this probability is $\pi/4$.
    }
    \label{fig:1}
\end{figure}
    
    Even in this case, the average energy per player increases with $ m $. However, comparing (18) and (21), we see that, as the number of players grows, the ratio between the energy of the negotiation and that of the ground state becomes larger, reaching the asymptotic limit $e$. It means that with more players, i.e. more constraints, the average benefit found by a \textit{superpartes} matchmaker is larger than that with $m$ players negotiating between themselves. 
    
    From these results, we also expect that, with more players, the probability of reaching the ground state through the negotiation process is lower, as shown in the inset of Fig. 3.
    
    It is remarkable that our negotiation model can be mapped to two well-known models of statistical mechanics: directed percolation and directed polymer \cite{roux1996extremal,halpin1995kinetic}. Indeed, the $m$ players and the $N$ alternatives can be arranged as a $m\times N$ matrix as shown in Fig. 4. The element $\epsilon_{ij}$ of this matrix corresponds to the energy of alternative $i$ for player $j$.
    
    let us examine the following transfer matrix relations:
    \begin{align*}
    \text{directed percolation:}\quad\;\;\;\;\, \eta_{\;\!i,t}&=\max(\eta_{\;\!i,t-1},\epsilon_{i,t}) \;,\\
    \text{directed polymer:}\quad \quad \,\,\;\; \eta_{\;\!i,t}&=\eta_{i,t-1}+\epsilon_{\;\!i,t} \;. \end{align*}
    Here, the time $t=1,...,m$ represents the players, and the space $i=1,...,N$ represents the alternatives. At each time step, the energy is updated following the relations above, up to $t=m$. At the end of the process, minimizing $\eta_{\;\!i,m}$ in the first relation gives the alternative $i$ corresponding to $f$ (i.e. the negotiation solution) and, in the second relation, it gives the ground state energy. Indeed, in the directed percolation, each value $ \eta_ {i, m} $ corresponds to the worst ranking of the alternative $ i $ in the preference-lists of the $ m $ players. Since players make their proposals starting from the top of their lists, the alternative $ i ^ * $, such that $ \eta_ {i ^ *, m} = \min_i(\eta_{\;\!i,m}) $, coincides with the first alternative to have been proposed by all players. As mentioned in section 2, this is precisely the condition for an alternative to be accepted by all. In the directed polymer, instead, $ \eta_ {i, m} $ is the sum of the rankings of the alternative $ i $ in the lists of the $ m $ players. Intuitively, the alternative $ i ^ * $, such that $ \eta_ {i ^ *, m} = \min_i(\eta_{\;\!i,m}) $, is the alternative that minimizes the sum of energies, i.e. the ground state solution. Hence, the two transfer matrix solutions correspond respectively to directed percolation and directed polymer in our context.
    
    \begin{figure}[!ht]
        \centering
        \includegraphics{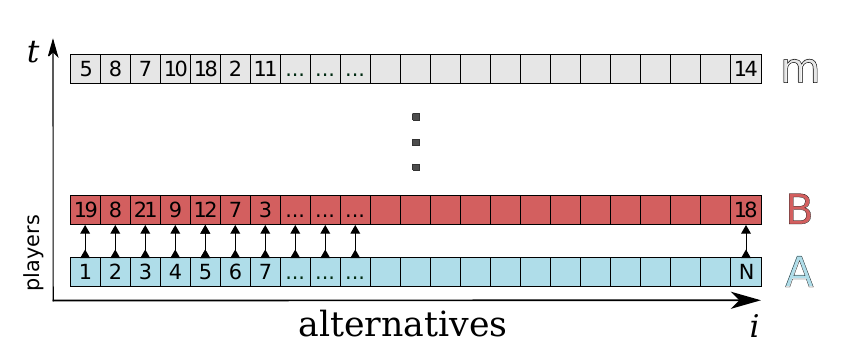}
        \caption{Transfer matrix of the negotiation problem: the rows represent players; the columns represent alternatives (ordered following A's preference-list). In the first row there are the energies assigned by player $ A $ to each of the $ N $ alternatives; in the second row there are the energies assigned by player $ B $; and so on up to player $ m $. In the transfer matrix, for each column, we iterate step by step from the bottom to the top. In this example, for the directed percolation, we have $ \eta_{\;\!1, A} = 1 $, and the first step gives $ \eta_{\;\!1, B} = \max (\eta_{\;\!1, A}, 19) = 19 $. By updating the energies step by step, up to $ t = m $, the agreed alternative can be found by minimization, i.e. $\min_i(\eta_{\;\!i,m})$
        }.
        \label{fig:my_label}
    \end{figure}
    
    Even though the negotiation problem seems to be hard, it is now mapped to well-known models that make it easily solvable, numerically, via the transfer matrix methods. However, our analytical solutions are still valuable as transfer matrix methods do not guarantee exact solutions. Combining our exact solutions and the transfer matrix approaches, we obtain both an analytical understanding and a fast algorithm to get numerical results. Moreover, the transfer matrix method has the advantage to yield all the degenerate solutions: each degenerate solution corresponds to a standard negotiation process with a different first-mover. Therefore, with the transfer matrix approach, one finds all the problem's outcome with a single sweep.
    
    Finally, we solve the ground state (21) by following a conventional statistical mechanics approach. Let us define the partition function of the system:
	\begin{equation}
	    Z_{\beta}=\sum_{i=1}^N e^{-\beta \epsilon_i} \;.
	\end{equation}
	Note that the partition function (22) is equivalent to that of the Random Energy Model (REM) by Derrida \cite{derrida1981random}.
	
	To find the free energy, we are interested in the average of the logarithm of the partition function over a distribution $\rho(\epsilon)$.
	
	We have: $F_\beta=\int\limits_0^\infty \frac{dl}{l} \left[e^{-l}-
	    \left(\int\limits_0^\infty d\epsilon \rho(\epsilon)e^{-l e^{-\beta \epsilon}}\right)^N\right]
	$.
	Here the variables $\epsilon_i$ follow the distribution $\rho(\epsilon)=\int dx_1..dx_m \delta\left(\epsilon-\sum_{i=1}^m x_i\right)$. Changing the variable to $l=e^{\beta u}$, and knowing that $\lim\limits_{\beta\rightarrow\infty}e^{e^{-\beta x}}=\theta(x)$, in the zero-temperature limit we obtain:  
    \begin{equation}
    \begin{split}
	    F_{\infty}=\int\limits_{0}^\infty du \left(
	    1-\int\limits_0^{u} d\epsilon \rho(\epsilon)
	    \right)^N = \int\limits_{0}^\infty du P(\epsilon_{{\scriptscriptstyle 0}}>u)=\overline{\epsilon_{{\scriptscriptstyle 0}}} \;,
	\end{split}
	\end{equation}
    that is the same ground state found in (21). 
    
    It is enlightening that, because of its simplicity, the traditional REM obtained with the (non-broken symmetry) Replica method can be reproduced by a procedural-combinatorial approach, making the negotiation problem particularly interesting. The latter method enjoys the advantage to be intuitive at each step, and the results can be verified numerically. On the other hand, the physical insights in the former method remain obscure.

    \section{Conclusion}

    
    To conclude, we introduced a simple model of negotiation, where different players with partially overlapping interests must find an agreement. We studied its analytical and numerical solutions with our new probabilistic procedure as well as with standard statistical mechanics approaches
    , showing that it can be mapped to a class of well-known physics problems.
    
    Moreover, it is an alternative model to those existing in the literature \cite{conc2,conc1,conc3,conc4}, which enriches the understanding of the negotiation process by proposing a different perspective. For example, it is possible to compare negotiation processes with bounded rational and hyper-rational players. In turn, this can help to find insights about people's behaviour and reasoning.
    
    We believe that the negotiation model is one of the simplest model dealing with such complex situations and, beyond its potential applications and extensions, it elucidates in a intuitive way the connection between social sciences and statistical mechanics.

    \section*{Acknowledgement}
    The authors would like to thank Fei Jing, Guiyuan Shi, and Ruijie Wu who supported this work with relevant discussions. This work was partially supported by the Swiss National Science Foundation (grant no. $200020\_182498/1$).


\appendix

 \bibliographystyle{elsarticle-num} 
 \bibliography{NPref}





\end{document}